\documentclass[aps,prl,superscriptaddress,
  rsi,amsmath,amssymb,
 reprint,
 floatfix,
]{revtex4-2}

\usepackage{graphicx}
\usepackage{dcolumn}
\usepackage{bm}
\usepackage{comment}
\usepackage{lipsum}
\usepackage{textcomp}
\usepackage{gensymb}
\usepackage{color,colordvi}
\usepackage[usenames,dvipsnames]{xcolor}
\usepackage{hyperref}

\usepackage{makecell}

\begin{document}

\title{Unexpected Behavior of Ultra-Low-Crosslinked Microgels in Crowded Conditions}

\author{Susana Mar\'in-Aguilar}
\email{susana.marinaguilar@uniroma1.it}
\affiliation{Department of Physics, Sapienza University of Rome, Piazzale Aldo Moro 2, 00185 Roma, Italy}
\author{Emanuela Zaccarelli}
\email{emanuela.zaccarelli@cnr.it}
\affiliation{CNR Institute of Complex Systems, Uos Sapienza, Piazzale Aldo Moro 2, 00185, Roma, Italy}
\affiliation{Department of Physics, Sapienza University of Rome, Piazzale Aldo Moro 2, 00185 Roma, Italy}

\date{\today}

\begin{abstract}
Ultra-low-crosslinked (ULC) microgels are among the softest colloidal particles nowadays routinely 
synthesized experimentally. Despite a growing literature of experimental results, their microscopic behavior under crowded conditions is yet to be revealed. To this aim, we resort to realistic monomer-resolved computer simulations to investigate their structural, mechanical, and dynamical properties across a wide range of packing fractions. Using particle-resolved analyses, we unveil the role of outer chains in the ULCs, which manifest in peculiar behaviors, utterly different from those of regularly crosslinked microgels. In particular, we report the absence of faceting and the dominance of interpenetration between microgels at high densities. Furthermore, we observe {a strong suppression of the structural reentrance characteristic of Hertzian-like particles, that is accompanied by the lack of a dynamical arrest transition, even well above random close packing. We further explore the change of behavior of the suspensions by lowering the crosslinker concentration and the single-particle density, providing strong evidence of the uniqueness of ULCs in the current landscape of microgels.} Altogether, our results establish ULCs as a distinct class of soft colloids in which polymeric degrees of freedom are highly predominant over colloidal ones, providing for the first time a robust, microscopic framework to interpret their unusual behavior.
\end{abstract}

\maketitle

\section*{Introduction}

Ultra-low-crosslinked microgels (ULCs) are a special class of colloidal polymer networks synthesized in the complete absence of any crosslinking agent~\cite{gao2003cross}. The preparation of Poly-N-isopropylacrylamide (PNIPAM) ULCs follows the same protocol as that of regular microgels, presenting no additional difficulties in their synthesis, and thus making them widely investigated by several experimental groups~\cite{bachman2015ultrasoft,virtanen2016persulfate,
scotti2019exploring,scotti2020flow,islam2021emergence,tennenbaum2021internal,burger2025suspensions}. The connectivity of the network arises from self-crosslinking between PNIPAM monomers during the polymerization process, occurring through chain transfer reaction. Recent scattering measurements of ULC form factors were compared to monomer-resolved simulations~\cite{hazra2023structure}, which allowed the estimation of the extremely small amount of these effective crosslinkers, between 0.1\% and 0.3\% molar fraction.

From Dynamic Light Scattering (DLS) measurements, it is well-established that ULCs behave rather similarly to crosslinked microgels with a similar volume phase transition temperature~\cite{hazra2023structure}. However, {their softness has been quantified by different experimental studies: their swelling ratio measured by DLS is unusually low~\cite{hazra2023structure,burger2025suspensions}, their elastic moduli appear to be more than one order of magnitude smaller than standard microgels~\cite{bachman2015ultrasoft,houston2022resolving,martinelli2025hierarchical}
and they display a much larger deformability at surfaces~\cite{bachman2015ultrasoft}.}
The intrinsic softness of ULCs also gives rise to a different phase behavior~\cite{scotti2020phase}, comprising stable bcc crystals, normally absent~\cite{paloli2013fluid} or just metastable~\cite{gasser2013transient} in crosslinked microgels.
Moreover, pioneering works at the boundary between bulk and interfacial behavior~\cite{scotti2019exploring, scotti2020flow} have revealed a transition from a colloidal-like to a polymer-like behavior, arising from the peculiar fluffy conformation of ULCs. Finally, a recent work pointed out the possible absence of a proper dynamical arrest transition in some ULCs~\cite{burger2025suspensions}, reinforcing the polymeric analogy at high densities.

Notwithstanding this vast body of experimental works, a detailed microscopic description of the structural and dynamical changes of ULCs
across varying packing fractions up to crowded conditions is still missing. While such insights have been provided by super-resolution microscopy measurements~\cite{conley2017jamming,conley2019relationship} and Molecular Dynamics (MD) simulations for realistic monomer-resolved 5\%-crosslinked microgels~\cite{del2024numerical,nikolov2020behavior}, similar results for ULCs have not yet been reported.

Given the low-density of the ULC network, a minimal number of monomers per particle is required in order to yield a `particle' with a significant amount of crosslinker. Therefore, in this work we perform extensive simulations of $N=54$ ULC microgels, each comprising $N_m\sim 16000$ monomers, {of which $0.2\%$ are crosslinkers. These ULC microgels are approximately three times larger than standard $5\%$-crosslinked microgels reported in previous simulations~\cite{del2024numerical}, which were carried out with $N_m\sim5000$ monomers, of which $\sim250$ were crosslinkers.}
{Notwithstanding this, here we investigate the suspension of ULCs in a wide range of generalized packing fractions $\zeta$, vastly exceeding random close packing (RCP), where $\zeta$ is the nominal packing fraction occupied by the microgels, computed using their volume in dilute conditions.}
This allows us to determine the interplay of different regimes such as deswelling, faceting and interpenetration\cite{lyon2012polymer} under crowded conditions.

Rather unexpectedly, our results provide profound differences between ULCs and 5\%-crosslinked microgels, starting from the fact that the former show a pronounced deswelling already well below random close packing, when microgels are not yet in direct contact (on average). This occurs due to the presence of long dangling chains in the periphery of the microgels, that occasionally meet and retract, yielding a vast reduction of the overall microgel size, even before microgels contact each other. This chain-dominated regime is reflected in most of the microscopic observables that we have studied, and has no counterpart in 5\%-crosslinked microgels. Given that these outer chains are mostly invisible experimentally, it is important that their features also manifest in other quantities. For example, they appear in the anisotropic shape of the microgels, which, contrary to expectations, tend to become more spherical, even excluding the presence of the outer chains. This is again at odds with standard microgels and leads us to exclude the onset of faceting in ULCs, a feature confirmed in the evolution of the system at high densities, directly visible in the simulation snapshots. In addition, the external chains give rise to a predominance of interpenetration among different microgels, to a much greater extent than what found for regular microgels. 

The consequences of these single-particle modifications are profound on the collective behavior, yielding {a significant suppression of the} reentrance for the radial distribution function. Indeed, its peak is found to remain very small in the whole investigated range of $\zeta$ and cannot be described by a Hertzian-like picture even for very low density. In addition, we do not observe the suspension to undergo a clear dynamical arrest, opposite to what observed for regular microgels. Both these features can be attributed to the prominent role of entanglements, as discussed in the following, which confirm the hypothesis of these microgels to become more and more similar to polymers with increasing crowding.

{To shed light on whether the differences between ULCs and 5\%-standard microgels arise from a continuous crossover or a qualitative change of behavior, we additionally perform extensive simulations of standard microgels in which we vary the key control parameters of their topology independently. Specifically, we simulate standard microgels with reduced crosslinker concentration down to 0.2\%, matching the ULC amount of self-crosslinks. Next, we address the role of the monomer density by simulating microgels with the same ULC density but higher crosslinker concentration. Finally, we examine the impact of network connectivity by considering ULCs with crosslinkers having a valence of 4. We thus report the behavior of the key observables for all these different microgels as a function of generalized packing fraction.}

{Altogether our results confirm the uniqueness of ULCs with respect to standard microgels, establishing them as a distinct class of soft colloids with peculiar structure and dynamics.} The present work is thus expected to stimulate new experiments able to discriminate these features in different laboratory samples at the microscopic scale. In addition, it will be important to properly locate ULCs on the softness axis~\cite{vlassopoulos2014tunable} with respect to other soft polymeric colloids, in order to assess their similarities and differences with respect to others, in particular with the recently proposed star-like microgels~\cite{ballin2025star}, also characterized by an extremely low softness, maybe even ultrasoft~\cite{likos2006soft}.

\section*{Results}
\begin{figure}[t]
\begin{center}
\includegraphics[width=0.8\linewidth]{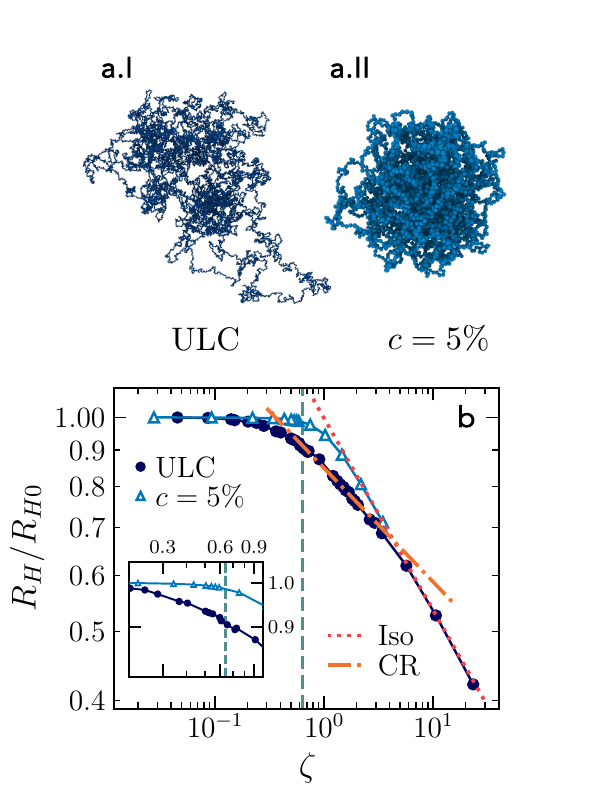}
\end{center}
\vspace{-0.5cm}
\caption{
\textbf{Deswelling behavior}  Snapshots of a.I) an ultra-low crosslinked microgel (ULC), and a.II) a microgel with a crosslinker concentration of $c=5\%$ at the lowest generalized packing fraction $\zeta$, respectively; 
(b) hydrodynamic radius $R_H$ normalized by its dilute limit $R_{H0}$ for ULC microgels (circles) and microgels with $c=5\%$ (triangles). The vertical dashed line denotes the random closed packing (RCP) value $\zeta_{\mathrm{RCP}}\sim 0.64$.  
The inset shows that ULCs begin shrinking already well below $\zeta_{\mathrm{RCP}}$, in a so-called chain regime (CR) dominated by their long-free chains. Here, the size reduction follows an apparent power law $R_H/R_{H0}\sim \zeta^{-1/6}$ (dash-dotted line). At high enough $\zeta$, deswelling becomes isotropic also for ULCs as well as for $c=5\%$ microgels, with $R/R_{H0}\approx \zeta^{-1/3}$, indicated by a dotted line.} 
\label{fig:packing}
\end{figure}

\subsection*{ULC vs. standard microgels with c=5\%} {We start by comparing results obtained for ULC microgels to available simulations of standard ones with 5\% crosslinker concentration, the only other case already reported in the literature so far~\cite{del2024numerical}.}

\textbf{Microgels deswelling} We begin by investigating the effects of crowding on the size of ultra-low crosslinked microgels (ULCs). Unlike regular microgels, ULCs are extremely soft, and can deviate significantly from a spherical shape. Representative snapshots of a ULC microgel and a regular one with crosslinker concentration of $c=5\%$ are reported in Figure~\ref{fig:packing}~a.I, a.II, respectively, to illustrate this contrast. While the ULC adopts an extended, loose conformation, the regular microgel remains rather compact. Consequently, when calculating the particle size as the cube root of the volume, as is commonly done,  the strong anisotropy of ULCs is neglected. Moreover, the presence of very long outer chains may lead to an overestimation of the particle volume. Despite these limitations, we still resort to the calculation of the hydrodynamic radius to quantify the size dependence of the microgels as a function of the nominal packing fraction $\zeta$ (see Methods Section, Equation~\ref{eq:rh} and Equation~\ref{eq:zeta}). This approach enables the comparison with available experimental data~\cite{scotti2019deswelling}. 

\begin{figure*}[th!]
\begin{center}
\includegraphics[width=0.99\linewidth]{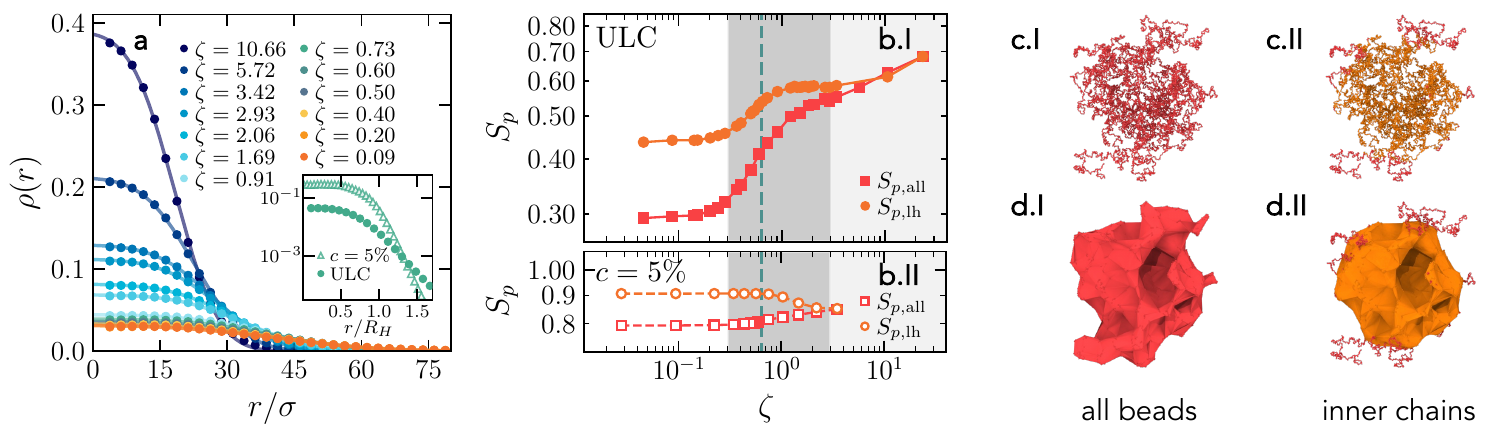}
\end{center}
\vspace{-0.5cm}
\caption{
\textbf{Shape evolution} a) Density profiles $\rho(r)$ for different values of $\zeta$. Lines correspond to fuzzy-sphere fittings. {The inset shows a comparison of normalized $\rho(r)$ between ULC (circles) and microgels with $c=5\%$ (triangles) at $\zeta\approx0.85$.} b) Shape anisotropy $S_p$ as a function of $\zeta$ for I) a system of ULC and II) microgels with $c=5\%$ accounting for different structural components, respectively. The ULC chain-dominated regime is shown in dark-gray, while the isotropic regime with light-gray. Dashed green line represent the random-closed packing of hard-spheres, $\zeta_{\mathrm{RCP}}=0.64$. c) Snapshot of a ULC at the lowest $\zeta$, showing I) all monomers and II) monomers with $|r_i-r_{\mathrm{cm}}| < l_h$. d) I-II corresponding surface meshes.  
}
\label{fig:anisotropic}
\end{figure*}

The hydrodynamic radius $R_H$, normalized by a reference value $R_{H0}$ obtained under dilute conditions, is shown as a function of $\zeta$ in Figure~\ref{fig:packing}b, together with the corresponding data for $c=5\%$ microgels, previously reported in Ref.~\cite{del2024numerical}. We first start by discussing the behavior for the crosslinked microgels. At low $\zeta$, where particles do not significantly interact, they maintain a constant $R_H$. Only when microgels come into contact, approximately around the hard-sphere random-closed packing fraction, $\zeta_{\mathrm{RCP}}\sim0.64$, the $c=5\%$ microgels start to shrink. For $\zeta \gtrsim 1.5 $, the scaling $R_H/R_{H0}\sim \zeta^{-1/3}$ emerges, consistent with uniform compression of spherical particles~\cite{del2024numerical}.

Here, we report a completely different behavior of ULC microgels, which begin to shrink much before reaching $\zeta_{\mathrm{RCP}}$, as evidenced in the inset of Figure~\ref{fig:packing}b, which shows that at this packing fraction ULCs have already decreased their size by about 10\%. For $\zeta > \zeta_{\mathrm{RCP}}$, two distinct shrinking regimes can be identified. At intermediate $\zeta$ values, ULCs shrink much more gradually, following an apparent power law $R_H/R_{H0}\sim \zeta^{-1/6}$. Then, for $\zeta\gtrsim 3.0$, they recover the isotropic shrinking behavior, merging with regular microgels. {The crossover value $\zeta$ is approximately identified by performing a two-power-law fit and selecting the value that minimizes the total squared error of the combined fit}. We speculate that the early shrinking of ULCs, followed by their more gradual deswelling at intermediate $\zeta$, are manifestations of the strong deformations of the outer shell, caused by the overlap of their very long outer chains, in a so-called chain-dominated regime (CR). Notably, at the same intermediate $\zeta$, the size reduction of the ULCs is always slightly larger than that of $c=5\%$ microgels, in qualitative agreement with the experimental results of Scotti et al.~\cite{scotti2019deswelling}.

\begin{figure}[h!]
\begin{center}
\includegraphics[width=1.0\linewidth]{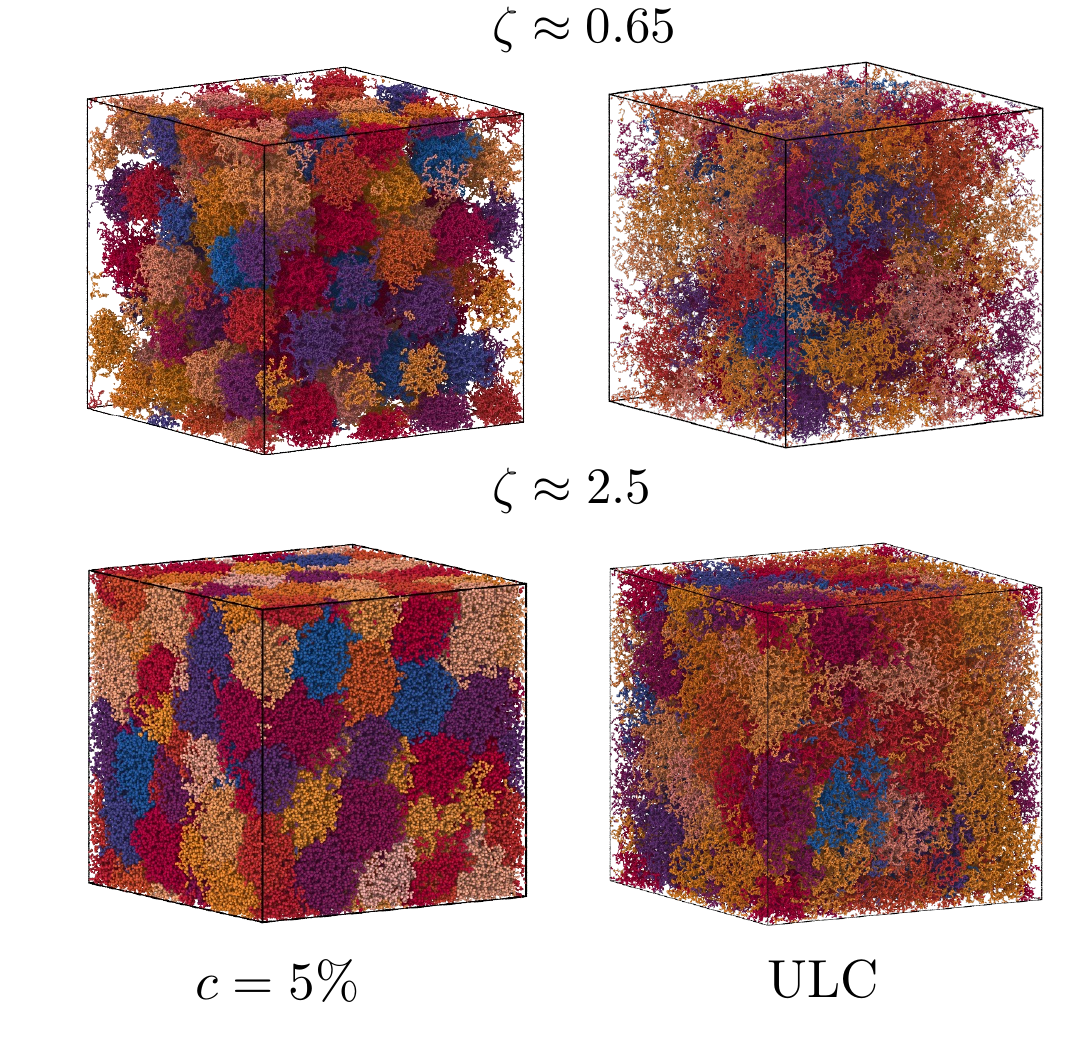}
\end{center}
\vspace{-0.5cm}
\caption{\textbf{Absence of faceting in ULCs} Representative snapshots of a system of a) microgels at $c=5\%$ and b) ULCs at a $\zeta\approx0.65$ (top) and $\zeta\approx2.5$ (bottom). }
\label{fig:manymgel}
\end{figure}
\textbf{Shape evolution}
Having quantified the ULC shrinkage behavior, we now turn to its shape evolution. Given the loose structure of an isolated ULC microgel, shown in Figure~\ref{fig:packing}~a.I, it is natural to expect that this will evolve and deform differently from $c=5\%$ microgels.

We first characterize the overall structural changes by computing the density profile $\rho(r)$ of ULCs
as a function of $\zeta$. The absence of a well-defined structure is immediately apparent from the $\rho(r)$, as shown in Figure~\ref{fig:anisotropic}a. In general, $\rho(r)$ displays a smooth shape, nearly Gaussian shape at all packing fractions, becoming only slightly more compact as $\zeta$ increases.  Although a fuzzy sphere fit (Eq.~\ref{eq:fuzzysphere}) is always found to reasonably describe the density profile, the so-called core is much smaller than the corona and remains indistinguishable in $\rho(r)$ at all studied $\zeta$. {Indeed, the density of ULCs in the center of the microgel is roughly one order of magnitude lower than that of $c=5\%$ microgels at a corresponding packing fraction, as shown in the inset of Fig.~\ref{fig:anisotropic}a. Consistently, ULCs exhibit a much more extended corona than standard microgels}.  This indicates that ULCs do not possess a well-defined core, unlike regular or star-like microgels~\cite{ballin2025star}, and are thus expected to exhibit markedly different behavior.

To gain a deeper insight into the shape of the microgels, we calculate the dimensionless anisotropy shape parameter, $S_p$, which measures the deviation from a perfect sphere (see Methods Section, Equation~\ref{eq:nu}), as a function of $\zeta$. By definition, $S_p = 1$ for spherical shapes, with lower values indicating progressively more anisotropic shapes. To ensure a more accurate description, we now rely on the computation of the ULCs' volume by constructing a surface mesh that encloses each particle. 

We begin by calculating $S_p$ accounting for all monomers, denoted as $S_{p,\mathrm{all}}$. These results are reported in Figure~\ref{fig:anisotropic}~a.I for ULCs and Figure~\ref{fig:anisotropic}~a.II for the previously reported $c=5\%$ microgels. 
As already shown in Ref.~\cite{del2024numerical}, microgels with $c=5\%$ show some deviations from spherical symmetry even under dilute conditions, as observed in Figure~\ref{fig:anisotropic}~a.II, with $S_{p,\mathrm{all}}\sim 0.8$.
However, as expected, ULCs exhibit a much more anisotropic shape even at the lowest $\zeta$, where $S_{p,\mathrm{all}}\sim 0.3$. This deviation is clearly visible from their structure and corresponding surface mesh, shown in Figure~\ref{fig:anisotropic}~b.I and c.I, respectively. These appear rough and irregular, thus yielding significantly lower values of $S_{p,\mathrm{all}}$, (filled square symbols in Figure~\ref{fig:anisotropic}~a.I) as compared to $c=5\%$ microgels (open square symbols in Figure~\ref{fig:anisotropic}~a.II). At low enough $\zeta$, the microgels retain their initial shape, with the $c=5\%$ microgels remaining essentially unperturbed up to $\zeta_{\mathrm{RCP}}$, while the ULCs undergoing significant shape deformation already for $\zeta\sim 0.3$. This packing fraction approximately marks the onset of the chain-dominated regime, discussed earlier in the context of the size behavior and highlighted in Figure~\ref{fig:packing}~b. Indeed, 
at such low $\zeta$, ULCs are not yet in direct contact with their neighbors, however, occasional encounters between their long outer chains cause them to deform from their initial state. Given that these chains will eventually turn inwards, the overall particle size decreases and the shape becomes more spherical. This trend is captured by the gradual increase of $S_{p,\mathrm{all}}$ with $\zeta$, which persists well beyond random close packing. 

Based on the dependence of $S_{p,\mathrm{all}}$ on $\zeta$ for ULCs, we can tentatively identify two distinct deformation regimes that are the direct counterparts of what already observed in the deswelling behavior: i) a rapid change at low/intermediate $\zeta$ ($\gtrsim0.3$) and ii) a slower deformation for  $\zeta\gtrsim 1.0$, that becomes more  
 similar to the behavior of the $c=5\%$ microgels. Nonetheless, even at the largest studied $\zeta$, ULCs remain very far from being spherical and well below the values obtained for regular microgels. 

To better quantify the role of the outer chains, we take a similar approach to the one proposed in Ref.~\cite{del2024numerical}. Specifically, we calculate $S_p$ after removing the contribution of such chains, that is defined as $S_{p,\mathrm{lh}}$. This is done by considering only monomers within $|r_i-r_{\mathrm{cm}}| < l_h$, with $l_h = R_c + \sigma_s$, where $R_c$ and $\sigma_s$ are the core radius and the width of the corona obtained from the fuzzy-sphere fits of the density profile under dilute conditions (see Methods). The selected inner chains and their surface mesh are shown in Figure~\ref{fig:anisotropic}~b.II–c.II. By construction, these correspond to monomers inside a sphere of radius $l_h$, for which one would expect $S_{\mathrm{p,lh}} \approx 1$. However, differently from c=5\% microgels  where $S_{\mathrm{p,lh}}$ increases and tends to the spherical limit (see open circles in Figure~\ref{fig:anisotropic}a.II) at low $\zeta$, for ULCs the value  of $S_{\mathrm{p,lh}}$ grows only moderately, remaining very far from unity due to the fluffy nature of the ultra-low-crosslinked network. 
Although $S_{\mathrm{p,lh}}$ is consistently larger than $S_{p,\mathrm{all}}$ under the same conditions, it only reaches $\approx0.4$ at low $\zeta$, with deformation still taking place well below $\zeta_{\mathrm{RCP}}$. At higher $\zeta$, the collapse of the outer chains is complete when $S_{p,\mathrm{lh}} = S_{p,\mathrm{all}}$, occurring around $\zeta \sim 1$, beyond which isotropic shrinking is also observed. 

\begin{figure}[th!]
\begin{center}
\includegraphics[width=0.8\linewidth]{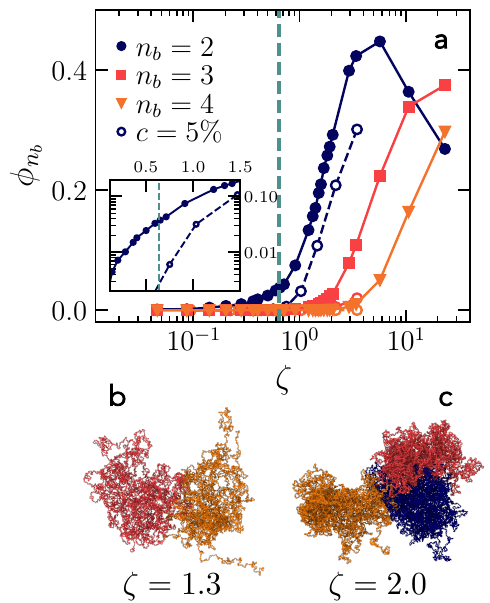}
\end{center}
\vspace{-0.5cm}
\caption{
\textbf{ Many-body overlaps} a) Many-body overlap volume fraction $\phi_{nb}$ as a function of generalized packing fraction $\zeta$ for ULCs, for $n_b=2$, $n_b=3$, and $n_b=4$. Open symbols correspond to the $n_b=2$ of microgels with crosslinker concentration $c=5\%$. The vertical dashed line indicates the random closed-packing fraction of hard spheres $\zeta_{\mathrm{RCP}}$. The inset reveals that, even at low values of $\zeta<\zeta_{\mathrm{RCP}}$, the ULCs already exhibit multiple overlaps.
b) Representative snapshots of two and three overlaps of ULCs at $\zeta=1.3$ and $\zeta=2.0$.}
\label{fig:overlap}
\end{figure}
A key difference between ULCs and c=5\% microgels arises from the behavior of $S_{p,\mathrm{lh}}$. While for ULCs this parameter increases at all $\zeta$, denoting the tendency of the particles to evolve toward a more regular shape, primarily driven by the collapse of the outer chains, the c=5\% microgels display an opposite trend with a reduction of $S_{p\mathrm{,lh}}$ above random close packing. This indicates an increasing deviation from spherical symmetry at high $\zeta$ and reflects the onset of faceting, consistent with super-resolution microscopy experiments on more highly crosslinked microgels~\cite{conley2019relationship}. Consequently, this phenomenon seems to be absent in ULCs. To confirm this result, we report representative snapshots in Figure~\ref{fig:manymgel} comparing c=5\% microgels with ULCs for two characteristic packing fractions, below and above $\zeta_\mathrm{RCP}$, i.e., $\zeta \sim 0.65$ and $\sim2.5$, respectively. The contrast between the two systems is striking, particularly at large $\zeta$, where the identity of the particles is clear for crosslinked microgels, while ULCs are much more heterogeneous and entangled, without displaying a clear particle structure. Altogether, this evidence strongly indicates that deformation and faceting are not dominant mechanisms for ULCs. Instead,  as previously shown in earlier works, their behavior resembles more that of a polymer suspension~\cite{scotti2019exploring,scotti2020flow}.
Moreover, the much stronger deformations observed in ULCs as compared to crosslinked microgels are consistent with the findings of Bachmann et al.~\cite{bachman2015ultrasoft}, who reported dramatic differences between the two systems based on their ability to spread on an interface.

\begin{figure*}[ht!]
\begin{center}
\includegraphics[width=1\linewidth]{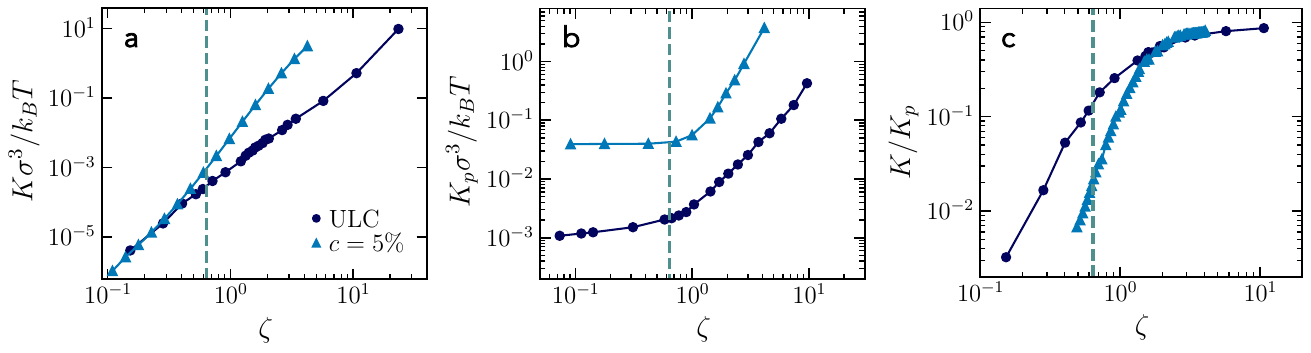}
\end{center}
\vspace{-0.5cm}
\caption{
\textbf{ Mechanical response} a) Bulk modulus $K$ as a function of generalized packing fraction $\zeta$ for a system of ULCs and $c=5\%$ microgels.
b) Single-particle bulk modulus $K_p$ obtained from spherical compression of isolated particles.
c) Ratio $K/K_p$ highlighting the connection between collective and single-particle responses. Dashed lines in all panels denote $\zeta_{\mathrm{RCP}}$.}
\label{fig:bulk}
\end{figure*}

\textbf{Many-body overlap}
The previous sections aimed to quantify two of the main mechanisms by which microgels readapt in response to external conditions: shrinking and deformation. However, there is also the possibility of interpenetration as a result of their polymeric nature. In order to estimate the importance of this scenario, we calculate the overlap volume fraction as $\phi_{nb}=V_{nb}/V$, where $V_{nb}$ is the volume occupied by $n_b=2,3, \ldots n$-microgels at the same time, as defined in the Methods Section. We show in Figure~\ref{fig:overlap}a the extent of these overlaps for the ULCs accounting for all beads and the corresponding pair overlaps ($n_b=2$) for the $c=5\%$ microgel from Ref.~\cite{del2024numerical}. ULCs display a sharp increase in two-body overlaps once the system enters the chain-dominated regime, well below RCP, as shown in the inset of Figure~\ref{fig:overlap}a.
The amount of overlaps is systematically larger than that observed for $c=5\%$ microgels at all comparable concentrations. In addition, significant contributions from three- and four-body overlaps are also present at higher $\zeta$, again in contrast to regular microgels~\cite{del2024numerical}. Representative configurations at $\zeta=1.3$ and $\zeta=2.0$, shown in Figure~\ref{fig:overlap}b, illustrate how overlaps first emerge from chain interdigitation and then evolve into collective, many-body contacts as density increases. Notably, the fact that the onset of two-body overlaps coincides with the chain-dominated regime ($\zeta\sim0.3$) identified in both the deswelling behavior (Figure~\ref{fig:packing}c) and the shape anisotropy (Figure~\ref{fig:anisotropic}a), suggests a consistent scenario where the increase of packing fraction soon drives ULCs into a regime dominated by shrinking and interpenetration. The latter persists up to very high $\zeta$, remaining always more pronounced than deformation, which actually tends to be reduced, giving rise to a completely different picture with respect to c=5\% microgels, where the latter is much more significant with respect to interpenetration~\cite{del2024numerical, conley2019relationship}.

\textbf{Mechanical response: Bulk modulus}
The structural changes observed in ULCs directly reflect their extreme softness, which, in turn, is expected to also strongly influence their mechanical response to compression. To probe this, we calculate the bulk modulus of the suspension, defined as $K=-V \partial P/\partial V$, as a function of $\zeta$, which is a measure of the collective mechanical response of the system, and compare it to that of the $c=5\%$ microgels.

At low $\zeta$, the overall behavior of ULC and $c=5\%$ microgel suspensions is nearly identical, as shown in Figure~\ref{fig:bulk}a. However, as crowding increases, the two systems depart from each other, with the bulk modulus of the $c=5\%$ microgels growing more rapidly, while that of the ULCs remaining smaller at comparable $\zeta$ values, as expected. Overall, at high $\zeta$, the bulk mechanical response of the ULC system is at least two orders of magnitude `softer' than that of their more crosslinked counterparts, consistent with experimental reports on ultrasoft microgels~\cite{bachman2015ultrasoft}.

To disentangle the role of single-particle elasticity from the collective response, we then compute the bulk modulus of an individual microgel, $K_p$, by performing spherical compressions of isolated ULCs (see Methods). Here, $K_p$ is obtained from the fluctuations of the surface-mesh enclosing the monomers, with the radius of the confining sphere mapped onto $\zeta$\cite{del2021two} (see Methods Section). This individual $K_p$ is shown in Figure~\ref{fig:bulk}b, alongside the corresponding for $c=5\%$ microgels. We find that $K_p$ exhibits a weak growth at low $\zeta$, followed by a steep increase starting approximately at $\zeta_{\mathrm{RCP}}$. Overall, the single-particle modulus of ULC microgels remain consistently lower than that of $c=5\%$ microgels by approximately two orders of magnitude across all studied $\zeta$ values. The most notable difference with respect to the crosslinked case is that $K_p$ is not constant at low $\zeta$, again confirming the influence of the outer chains in this regime.  

It is now interesting to compare the present numerical findings with experiments performed by Houston et al.~\cite{houston2022resolving}, who measured $K_p$ by applying osmotic pressure to dilute suspensions of microgels of varying $c$ by adding polymers that were rendered invisible to neutron scattering through selective deuteration. Their results revealed a difference of about two orders of magnitude in the bulk moduli between ULCs and more highly crosslinked microgels, in good agreement with the present results.
 
To connect single-particle and collective behavior, we then examine the ratio $K/K_p$, reported in Figure~\ref{fig:bulk}c. At intermediate and high $\zeta$ values ULCs and $c=5\%$ microgels display very similar trends. In this regime, the ratio for both systems approaches unity, indicating that, regardless of crosslinker concentration, the suspension’s mechanical response is dominated by single-particle elasticity, under very dense conditions. Indeed, at low $\zeta$ where particles are not in contact, the bulk modulus of the suspension is consistently smaller than that of a single microgel, reflecting that the overall system is easier to compress while particles are not in contact. Instead, when microgels become strongly interacting, the bulk mechanical response directly relies on the ability of individual particles to compress. As a consequence, the collective response is governed by $K_p$, in agreement with previous numerical and experimental studies~\cite{lietor2011bulk,nikolov2020behavior,del2024numerical}. Interestingly, this property, i.e., $K/K_p\rightarrow 1$ at high $\zeta$, seems to be a generic feature of all bulk suspensions of microgel particles, independently of $c$.

\begin{figure*}[t!]
\begin{center}
\includegraphics[width=1.0\linewidth]{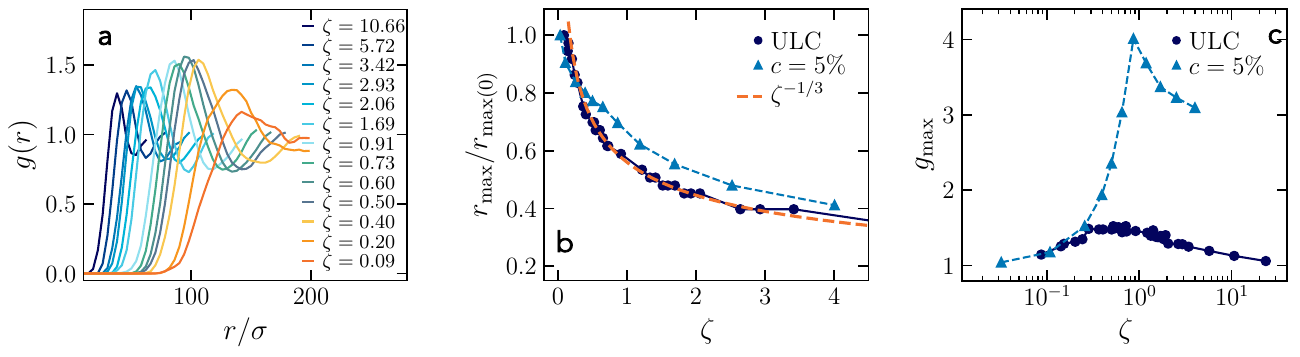}
\end{center}
\vspace{-0.5cm}
\caption{\textbf{Bulk structure} a) Radial distribution function $g(r)$ for ULC at different generalized volume fractions $\zeta$.  b) Normalized position of the first peak of the $g(r)$ for ULC and microgels with $c=5\%$; c) corresponding height of the first peak.
 }
\label{fig:gmax}
\end{figure*}

\textbf{Collective behavior}
Once established the extreme soft nature of the ULC and its consequences on their mechanical response, we turn our attention to its effect on their structural collective behavior. For regular microgels, it is well-known that at high concentrations a structural reentrance takes place~\cite{zhang2009thermal,paloli2013fluid}, manifesting as a non-monotonic behavior of the height of the main peak of the radial distribution function with increasing $\zeta$, as also confirmed in recent simulations~\cite{del2024numerical}.  However, the situation is rather different for ULC microgels.

To prove this, we calculate the $g(r)$ of the suspension over the entire range of $\zeta$ as reported in Figure~\ref{fig:gmax}a, and extract both the height $g_\mathrm{max}$ and the position of the first peak $r_\mathrm{max}$. Interestingly, the latter follows the expected decrease with packing fraction of homegeneous systems, i.e., $r_\mathrm{max} \sim \zeta^{-1/3}$, across the entire range of $\zeta$ as shown in Figure~\ref{fig:gmax}b. In contrast, the height of the peak $g_\mathrm{max}$ shows no significant increase with $\zeta$, as reported in Figure~\ref{fig:gmax}c. This indicates the absence {of almost any} local structural ordering at higher concentrations. This behavior is in stark contrast to the $g_\mathrm{max}$ for c=5\% microgels, from Ref.~\cite{del2024numerical}, also reported in the Figure. 

\begin{figure}[t!]
\begin{center}
\includegraphics[width=0.8\linewidth]{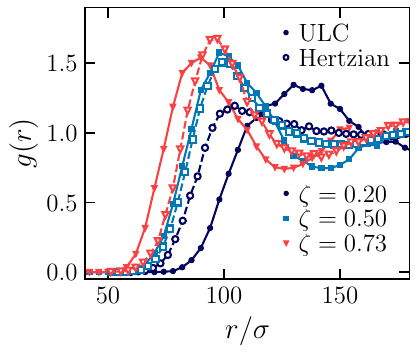}
\end{center}
\vspace{-0.5cm}
\caption{\textbf{Non-Hertzian behavior} Radial distribution function calculated from simulations of ULCs (full symbols) and obtained from an effective system interacting through a Hertzian potential at same $\zeta$ (open symbols) at three representative packing fractions. The Hertzian model is chosen to best match the numerical $g(r)$ at $\zeta=0.50$, amounting to a Hertzian strength at contact $\sim 50k_BT$ and to an effective size $\sigma_H\sim 105\sigma$.
 }
\label{fig:hertz}
\end{figure}

Another important difference with respect to crosslinked microgels is that in the low-$\zeta$ regime their behavior could be well-reproduced by using an effective Hertzian potential~\cite{paloli2013fluid,del2024numerical}. This does not hold for ULC microgels, as shown in Figure~\ref{fig:hertz}, where we report the results of the numerical $g(r)$ for ULCs in comparison with those obtained from a Hertzian potential, chosen to  describe the numerical behavior  at $\zeta=0.50$. While the agreement for this particular $\zeta$ can be considered satisfactory, it is clear that moving to another packing fraction, either larger or smaller, the effective potential does not capture the evolution of the $g(r)$ at all. Particularly, the evolution of the peak position is not compatible with a Hertzian shape of the potential. Similar unsatisfactory results are also found when using a harmonic or a gaussian model. Given that such a coarse-grained approach can only work in dilute conditions, we thus conclude that the failure of a Hertzian-like description is yet another manifestation of the dominance of the outer chains for ULC microgels under dilute conditions. As a consequence, it is not possible to rationalize the numerical findings in terms of effective spheres with some underlying fixed potential.

\begin{figure*}[t!]
\begin{center}
\includegraphics[width=1\linewidth]{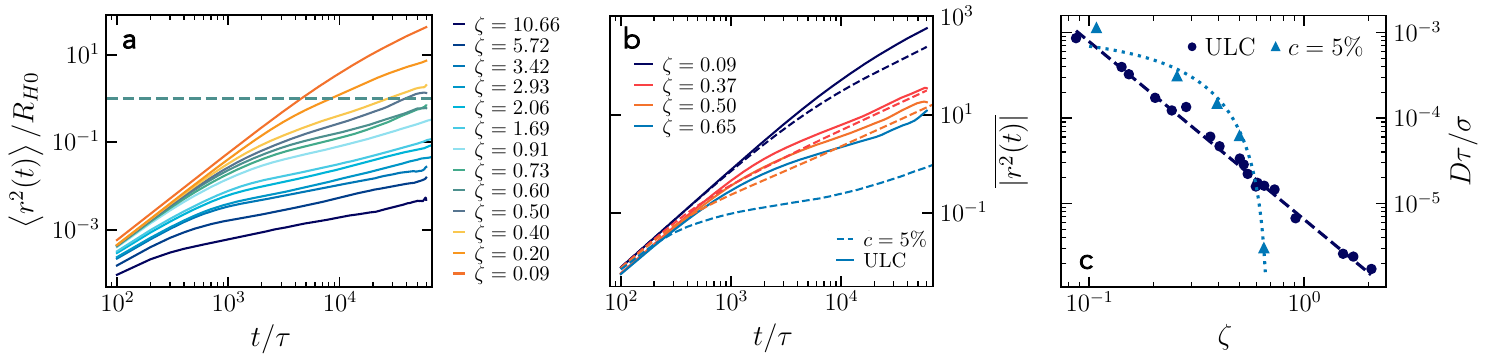}
\end{center}
\vspace{-0.5cm}
\caption{
\textbf{Dynamical behavior} a) Mean-square displacement (MSD) normalized with $R_{H0}$ for ULCs at different generalized volume fractions $\zeta$. Dashed lines correspond to one $R_{H0}$. b) Normalized MSD for ULCs (solid lines) and microgels with $c=5\%$ (dashed lines).
c) Long-time diffusion coefficient $D$ as a function of $\zeta$ for ULCs (circles) and $c=5\%$ microgels (triangles). Data for ULCs are well-fitted by a simple power-law $\sim \zeta^{-1.9}$ (dashed line), showing no apparent divergence at any give $\zeta$. Instead, c=5\% microgels are roughly compatible with a MCT arrest at $\zeta_{MCT}\sim 0.69$ and power-law exponent $\gamma\sim 2$ (dotted line).
}

\label{fig:dynamics}
\end{figure*}

 The anomalous, almost unstructured conformation of ULC suspensions has also  profound consequences on its dynamics. To investigate this aspect, we report the mean-squared displacement (MSD) of the center of mass of ULC microgels at different $\zeta$ in Figure~\ref{fig:dynamics}a. We observe no evidence of a dynamical reentrance, long ago predicted for Hertzian potential~\cite{berthier2010increasing}, in analogy to what already observed for standard microgels~\cite{del2024numerical,philippe2018glass}. This means that the dynamics becomes monotonically slower as $\zeta$ increases. Nevertheless, at high $\zeta$, ULCs exhibit significant larger displacements than regular microgels, reflecting the low energetic penalty for particle overlap. This is highlighted in Figure~\ref{fig:dynamics}b, where normalized MSDs for ULCs (solid lines) are compared to those of regular microgels (dashed lines) at a comparable $\zeta$.

To quantify this effect, we calculate the self-diffusion coefficient $D$ extracted from the long-time behavior of the MSD, that is  reported in Figure~\ref{fig:dynamics}c for both systems, but only for those state points where a diffusive behavior $MSD \sim t$ is recovered at long times. We find that for both types of microgels diffusion decreases steadily with increasing $\zeta$, confirming the absence of a reentrant behavior. Yet, while 5\%-microgels clearly approach dynamical arrest already for $\zeta \lesssim 1.0$, ULCs remain diffusive up to at least $\zeta \sim 2$. Above this packing fraction, we cannot reliably estimate $D$, because the MSDs shows sub-diffusive behavior with time, as evident from Figure~\ref{fig:dynamics}a. 
The dependence of the self-diffusion coefficient on $\zeta$ for regular microgels appears to be compatible with the behavior predicted by Mode-Coupling Theory~\cite{gotze1999recent}, as shown in Figure~\ref{fig:dynamics}c, approaching dynamical arrest at $\zeta_{MCT}\sim 0.69$, as evidenced by the sharp decrease of $D$ by at least two orders of magnitude.  In contrast, the ULC diffusion coefficient decreases in a much slower fashion, that is compatible with a divergence-free power-law dependence (dashed line in Figure~\ref{fig:dynamics}c) with a very small exponent $\sim -1.9$, which seems to indicate no proximity to a dynamical arrest.  

\subsection*{{Are ULCs really special?}} 
{All of the results presented so far suggest a strong dichotomy between ULCs and 5\% regular microgels. This observation raises the question of whether the ULC behavior reflects a true qualitative change, or if it can instead be continuously reached by tuning experimentally accessible parameters merging the two regimes.} {To address this question, we have performed additional simulations to bridge the gap between the two types of microgels, which differ not only in crosslinker concentration but also in the number density of the single microgel, with $\rho_{\mathrm{ULC}}$ being about a factor $\sim 3$ smaller than the one of standard microgels $\rho_{\mathrm{st}}$. In addition, due to their different chemistry, ULCs have a valence of (self-)crosslinking fixed to 3 rather than four. Therefore, to reveal the independent role of these parameters, we vary each of them separately. A summary of all performed simulations with their main behaviors is presented in Table~\ref{tab:simulations}. }

\begin{figure*}[t!]
\begin{center}
\includegraphics[width=1\linewidth]{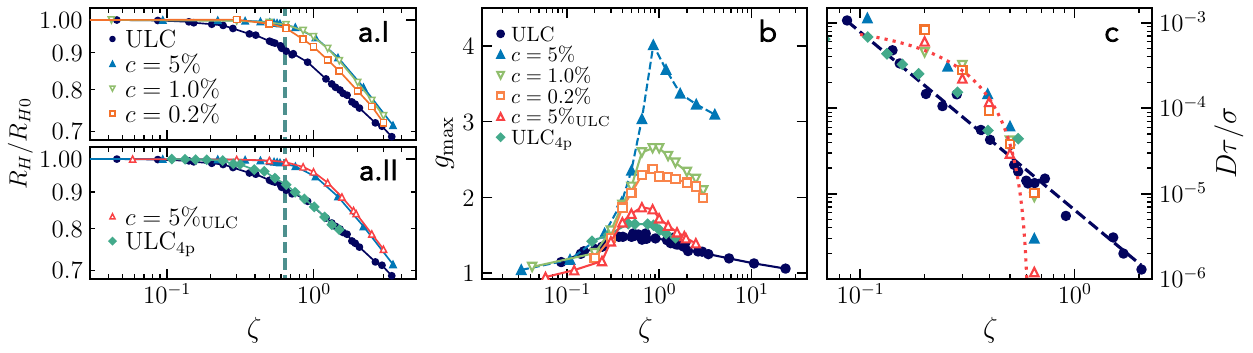}
\end{center}
\vspace{-0.5cm}
\caption{
\textbf{Varying density and crosslinker properties} {Deswelling behavior: $R_H$/$R_{H0}$ for a system of ULC$_{\mathrm{3p}}$ microgels (circle), standard-$c=5\%$ (filled triangles), (a.I) $c=1.0\%$ (inverted triangles) and $c=0.2\%$ (squares), the latter at fixed single-particle density $\rho_{\mathrm{st}}$, (a.II) ULC$_{\mathrm{4p}}$ microgels (empty triangles) and  $c=5\%_{\mathrm{ULC}}$ (empty triangles) both with  $\rho_{\mathrm{ULC}}$. b) Height of the first peak of the radial distribution function and c) corresponding long-time diffusion coefficient $D$ as a function of $\zeta$ for all systems.  The dotted line indicates a power-law dependence similar to standard  $c=5\%$ microgels in Fig.~\ref{fig:dynamics}(c)  for  $c=5\%_{\mathrm{ULC}}$, roughly describing also $c=1.0\%$ and $c=0.2\%$ systems.}
}
\label{fig:transition}
\end{figure*}

\textbf{Effect of crosslinker valence} {To isolate the role of crosslinker valence, we study ULCs assembled with crosslinkers of valence 4 (hereafter denoted ULC$_{\mathrm{4p}}$), as in standard microgels, while keeping all other parameters unchanged, including $\rho_{\mathrm{ULC}}$. Interestingly, we find that the behavior of these `more connected' ULCs is almost identical to that of the original ones. This is observed from the deswelling behavior, the height of the first peak of the $g(r)$ and the long-time diffusion coefficient, shown in Fig.~\ref{fig:transition}a-c, respectively, as a function of $\zeta$. These results demonstrate that a slight increase in crosslinker valence does not qualitatively change the polymer-like behavior characteristics of ULCs and their reduced tendency toward dynamical arrest.}

\textbf{{Effect of crosslinker concentration at fixed density $\rho_{\mathrm{st}}$}} {We then investigate the effect of $\%c$ in standard microgels. In particular, we aim to determine whether lowering $\%c$ to the amount corresponding to the ULCs, while keeping the standard microgel density ($\rho_{\mathrm{st}}$) fixed,  the peculiar behavior of ULCs can be recovered. We thus investigate standard microgels with $c=1\%$ and $c=0.2\%$ and also report the same key observables associated with the chain-dominated regime in Fig.~\ref{fig:transition}a-c for different generalized packing fractions.  It is evident that both microgels with low-crosslinker concentration behave similarly to $c=5\%$ standard ones. In particular, they exhibit isotropic shrinking immediately beyond the random close packing, with no signature of the chain-dominated regime characteristic of the ULCs. Structurally, the presence of a  reentrance is observed, with a reduced peak height as compared to the $c=5\%$ case. Finally, their diffusion coefficient also resembles that of $c=5\%$ microgels, including a similar dynamical arrest. These findings indicate that, lowering the crosslinker concentration alone, without changing the monomer density within the network, is not sufficient to bridge the gap between ULCs and standard microgels.}

\textbf{{Effect of monomer density}} {Finally, we examine the effect of reducing the single-particle density to the one of the ULCs, $\rho_{\mathrm{ULC}}$. To this end, we prepare microgels with $c=5\%$ but with a particle density matched to that of the ULC, which we denote as $c=5\%_{\mathrm{ULC}}$. Again, their behavior as a function of $\zeta$ is reported in Fig.~\ref{fig:transition}(a-c), together with all other microgels. Notably, we find that the $c=5\%_{\mathrm{ULC}}$ exhibits the same isotropic shrinking behavior as the standard $c=5\%$ microgels. Structurally, they display a mild reentrant behavior, closer to that of ULCs, than all standard microgels, signaling that the density is the dominant parameter for the collective arrangement of the microgels. However, turning to dynamical arrest, these microgels display a behavior more similar to standard microgels than to ULCs, with a transition to a glassy state.}

\begin{figure}[h!]
\begin{center}
\includegraphics[width=1\linewidth]{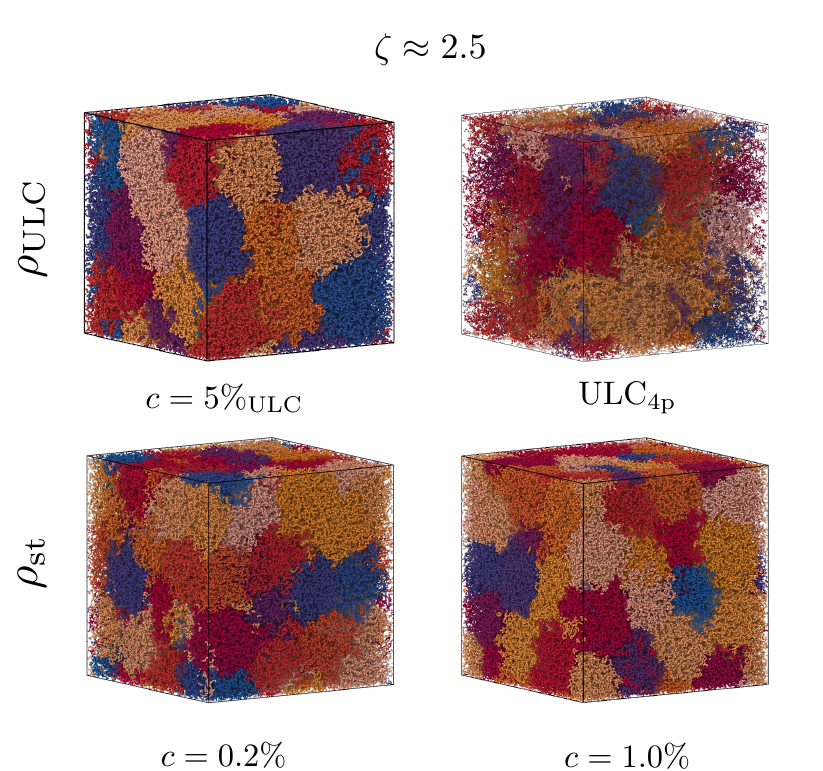}
\end{center}
\vspace{-0.5cm}
\caption{{Representative snapshots of the different microgels at a generalized packing fraction $\zeta\approx2.5$:  (top) $c=5\%$ with $\rho_{\mathrm{ULC}}$ and ULCs with a crosslinker valence of $4$; (bottom) standard microgels with $c=0.2\%$ and $c=1.0\%$ with particle density $\rho_{\mathrm{st}}$.} 
}
\label{fig:snap}
\end{figure}

{These results are confirmed by direct inspection of the configurations at high $\zeta$, shown in Fig.~\ref{fig:snap}. Indeed, it is evident that only ULC$_{\mathrm{4p}}$ microgels show a diffused polymeric architecture, while all the others clearly display the particles getting deformed and faceted, retaining their identity of colloidal particles.}

{Altogether, the investigation of additional microgel types demonstrates that the 
variation of only one of these control parameters does not allow to recover the 
peculiar behavior of ULCs. Indeed, reducing $\%c$ at standard particle density is not enough to reach the ULC regime for any of the three reported observables. Conversely, reducing the particle density at high crosslinker concentration, the reduction of the $g(r)$ is significant, but chain-dominated regime in deswelling is absent and dynamical arrest is retained. Hence, it is only the combined effect of low crosslinker concentration and low single-particle density that makes the ULC behavior truly special.
}

\begin{table*}[t]
\centering
\caption{Summary of all the simulated microgel systems, reporting the system label,  crosslinker valence, crosslinker concentration $\%c$, and the monomer density, with $\rho_{\mathrm{ULC}}\approx0.029$ and $\rho_{\mathrm{st}}\approx0.08$ corresponding to ULC and standard microgels, respectively. In addition, the main observed behaviors, namely, the presence of a chain-dominated (CR) regime, structural reentrance, and dynamical arrest, are schematized.}
\renewcommand{\arraystretch}{1.}
\setlength{\tabcolsep}{8pt} 

\begin{tabular}{|>{\centering\arraybackslash}m{1.8cm}
                |>{\centering\arraybackslash}m{2.0cm}
                |>{\centering\arraybackslash}m{0.6cm}
                |>{\centering\arraybackslash}m{1.8cm}
                |>{\centering\arraybackslash}m{1.6cm}
                |>{\centering\arraybackslash}m{1.8cm}
                |>{\centering\arraybackslash}m{1.6cm}|}
\hline
\makecell{\textbf{System}} &

\makecell{\textbf{Crosslinker}\\\textbf{valence}} &
\makecell{\textbf{\%c}} &
\makecell{\textbf{Monomer}\\\textbf{density} \(\rho\)} &
\makecell{\textbf{CR}\\\textbf{regime}} &
\makecell{\textbf{Structural}\\\textbf{reentrance}} &
\makecell{\textbf{Dynamical}\\\textbf{arrest}} \\
\hline\hline

ULC${_{\mathrm{4p}}}$ & $4$ & $0.2$ & $\rho_{\mathrm{ULC}}$ & yes  & suppressed & no \\
 ULC${_{\mathrm{3p}}}$ & $3$ & $0.2$ & $\rho_{\mathrm{ULC}}$ & yes  & suppressed & no \\
 $c=0.2\%$ & $4$ & $0.2$ & $\rho_{\mathrm{st}}$ & no  & yes & yes \\
  $c=1\%$ & $4$ & $1$ & $\rho_{\mathrm{st}}$ & no  & yes & yes \\
 $c=5\%$ & $4$ & $5$ & $\rho_{\mathrm{st}}$ & no  & yes & yes \\
  $c=5\%_{\mathrm{ULC}}$ & $4$ & $5$ & $\rho_{\mathrm{ULC}}$ & no  & yes/mild & yes \\
\hline
\end{tabular}
\label{tab:simulations}
\end{table*}

\section*{Discussion}
In this work, we have carried out an extensive exploration by molecular dynamics (MD) simulations of monomer-resolved ultra-low-crosslinked microgels in good solvent up to very crowded conditions. By systematically increasing the packing fraction, we have revealed how the loosely connected architecture of ULCs leads to structural, mechanical, and dynamical responses and compared to available results for standard microgels with higher crosslinker concentration.

First of all, it is important to remind ourselves that standard microgels at high concentration can exhibit three main types of response: deswelling, deformation (or faceting) and interpenetration~\cite{lyon2012polymer}. The interplay of these three regimes is subtle and was recently established to take place in this order through careful super-resolution microscopy experiments~\cite{conley2019relationship} and MD simulations~\cite{del2024numerical}. For ULCs, there are no available super-resolution experiments to date, thus we have to rely on the simulations for the moment, having previously established that our monomer-resolved model can realistically represent the internal structure of a single ULC microgel in bulk~\cite{hazra2023structure} and at interfaces~\cite{bochenek2022situ,gerelli2024softness}.

{Focusing first on the comparison with high-$c$ standard microgels}, the present MD results show two peculiar findings that are unique to ULCs. First of all, we find the onset of a chain-dominated regime that takes place at low and intermediate packing fractions, already emerging well below RCP. In this regime, the long outer chains of distinct ULC microgels encounter, so that they bend inwards, giving rise to a strong size reduction even when particles are not in direct contact. From our results, this regime is not only directly evident in the deswelling behavior, but also manifests in most of the observables examined in this work.

The other important finding is the absence of the faceting regime that is commonly attributed to microgels. Being ULCs largely anisotropic and fluffy under dilute conditions, they never get a proper convex shape (see Figure~\ref{fig:anisotropic}) so that they remain far from compact objects at all investigated densities. This is clearly visible in the high-$\zeta$ snapshots of Figure~\ref{fig:manymgel}, which again contrast ULCs with respect 
to standard microgels.

While faceting is missing, the open architecture of ULCs largely favours interpenetration. 
We quantified this phenomenon by calculating many-body overlaps and showing that two-body interpenetration again appears below RCP and that three- and four-body overlaps become increasingly relevant at large crowding. These findings overall confirm that the loosely-connected ULC architecture strongly promotes interpenetration and that this mechanism is at play at all packing fractions, differently from what observed for $c=5\%$ microgels, where this phenomenology was found to be rather marginal in a limited interval of $\zeta$.
These structural changes directly influence the mechanics of ULCs, that was addressed by calculating their bulk modulus. In particular, we found that the bulk modulus is roughly two orders of magnitude smaller than that of c=5\% microgels in agreement with recent experiments~\cite{houston2022resolving}.

In addition, we investigated the dynamical and structural collective behavior of the ULCs. {Unlike standard microgels, the structural reentrance of the radial distribution function is strongly suppressed for ULCs.} This is a rather unexpected result, since in the literature this feature is commonly attributed to softness of Hertzian-like systems. Then, why do extremely soft particles such as ULCs not show it? The reason is that they are so soft, that they fail to order even locally. Indeed, their extremely loose network connectivity and fuzziness allow monomers from different particles to interpenetrate almost freely, without generating the same type of elastic resistance observed in standard microgels. As a result,  the peak of the radial distribution function always remains very small, slightly exceeding one at all investigated packing fractions. This behavior turns out to be not at all compatible with a Hertzian model. Even trying out different effective models, we find that no simple effective pair potential can describe ULCs collective structure.

Another important result of our simulations is that ULCs remain diffusive even well above concentrations where standard microgels undergo arrest, i.e., above RCP. The evolution of the mean-squared displacement calculated in this work does not allow us to estimate an arrest packing fraction for ULCs.
{Interestingly, these results are consistent with very recent measurements of considerably small ULC microgels, for which the relaxation time measure by Dynamic Light Scattering is compatible with an exponential growth, while the viscosity extracted from rheological measurements shows no evidence of glass transition up to at least $\zeta\sim 3.0$~\cite{burger2025suspensions}. This behavior contrasts earlier experimental findings  for slightly larger ULC microgels~\cite{scotti2020flow} and also more recent, extended studies on similarly -sized microgels, performed by Martinelli and coworkers~\cite{martinelli2025hierarchical}, both observing dynamical arrest for $\zeta$ above 1.0. 
\\
To understand these results, it thus seems to be important to correctly take into account the internal structure of these soft particles, in order to determine their dynamics. Indeed, the samples by Burger and coworkers~\cite{burger2025suspensions} seems to be incompatible with the fuzzy sphere model observed for previously studied ULCs~\cite{scotti2019deswelling} and more similar to star-like microgels~\cite{ballin2025star}. Furthermore, we remark that the present computational model for ULCs was calibrated through a quantitative comparison with form factors across temperatures for very large ULC microgels~\cite{hazra2023structure}, for which no surfactant was added to the synthesis. The presence of contrasting experimental results seems to suggest that perhaps some aspects of the chemical synthesis of the microgels, particularly the addition of surfactant, may well alter the final particle properties, making them more locally dense. Indeed, by varying the internal microgel density, we showed that some of the key properties analyzed in this work do show some variation. While the so-called $c=5\%_{\mathrm{ULC}}$ microgels are at present just a computational model, that we exploited only to disentangle the different control parameters, it may well be that some experimental samples show a variation in $c$
and in the internal $\rho$, to make the different properties among ULCs and standard microgels more continuous. In the same spirit, we also explored standard microgels with an unusually low $c$, i.e. the same  one as estimated for ULCs, which to our knowledge have never been explored in experiments. However, it appears that the addition of crosslinkers, introducing a different kinetic polymerization rate in the nucleation of the microgels, automatically sets a larger monomer density, which always makes the microgels synthesized in the presence of BIS in the `colloidal' category and utterly distinct from ULCs, as also shown by the present simulations. Therefore, we hope that the present work will further stimulate detailed comparisons between experiments and numerical simulations, assessing in detail the internal structure of the different ULC samples used by the various groups in order to clarify the experimental situation and to be able to decipher the ingredients leading to one behavior with respect to another.
Indeed, as shown in Fig.~\ref{fig:transition}, subtle differences in the structure can be relevant for the high density behavior of these very soft objects. It thus remains important to clarify the relationship between softness and rheology, as already pointed out in the pioneering review article by Vlassopoulos and Cloitre~\cite{vlassopoulos2014tunable}, and to carefully quantify their `softness' in the context of other existing soft particles~\cite{scotti2022softness}.
\\
In summary, the present results establish ULCs as a distinct class of microgels, with properties that are profoundly different from standard crosslinked ones, in agreement with experimental observations~\cite{bachman2015ultrasoft,scotti2019deswelling}. This is due to the unique combination of the lowering of the crosslinker concentration and of the unusually low internal particle density and we further showed that their peculiar behavior cannot result from varying each parameter separately. In physical terms, this can be explained by the fact that their behavior is largely dominated by polymeric features, including the chain regime and the strong interpenetration, disfavouring mechanisms more typical of colloidal behavior, such as faceting and conventional glassy arrest. We speculate that the present findings may also provide a clear microscopic picture of the mechanisms by which the colloid-to-polymer transition occurs~\cite{scotti2019exploring,scotti2020flow}, so far elusive in bulk experiments and only detected at interfaces.}

Finally, the present findings pave the way for future studies aimed to clarify the role of softness in a wide variety of contexts, from the rheological response under external fields to the behavior in mixtures with other soft or hard colloids as well as their relevance as model systems for crowding in biological environments~\cite{saxena2014microgel,gaines2016adhesion}. More broadly, the unique combination of softness and interpenetrability positions ULCs as promising building blocks for the design of new classes of exceptionally soft, adaptive materials~\cite{brown2014ultrasoft} and for novel uses in biological contexts~\cite{babenyshev2025size}.

\subsection*{Simulations}
To explore the structural and dynamical features of ULCs under crowded conditions, we perform extensive molecular dynamics (MD) simulations in the $NVT$ ensemble, under various density conditions. 
The microgels are generated following the {\it in silico} synthesis protocol established in Refs.~\cite{gnan2017silico} that was adapted and validated for ULC microgels in Ref.~\cite{hazra2023structure}. In this approach, a standard microgel made of NIPAM and BIS crosslinkers is assembled using a binary mixture of patchy particles of radius $\sigma$ with two and four patches, corresponding to monomers and crosslinkers respectively. 
However, for ULC microgels there are no added crosslinkers and the network originates from self-crosslinking of NIPAM itself, which form bonds with valence three. We thus assemble microgels with $\sim 21000$ monomers each, of which $0.2\%$ have valence three, in a spherical confinement of radius $Z=55.5\sigma$, to maintain the low number density, $\rho_\mathrm{ULC}\approx0.029$, established for ULC microgels~\cite{hazra2023structure}.
At the end of the assembly process, we are left with a large network of $N_m=16255$ particles, due to the reduced connectivity of the system. Indeed, the average valence of the patchy binary mixture is only $2.002$. The remaining particles form isolated chains and are thus discarded from the subsequent analyses.

Once the microgel is formed, its topology is fixed by replacing the patchy interactions with the Kremer-Grest potential~\cite{kremer1990dynamics}. Hence, all particles interact with the Weeks-Chandler-Andersen (WCA) potential, 
\begin{equation}
    U_{WCA}(r)=\begin{cases} 4\epsilon\left[ \left(\frac{\sigma}{r}\right)^{12}-\left(\frac{\sigma}{r}\right)^6 \right]+\epsilon & r\leq 2^{1/6}\sigma \\
    0 & r> 2^{1/6}\sigma,
    \end{cases}
\end{equation}
where $r$ is the distance between two beads and $\epsilon$ controls the energy scale and corresponds to the unit of energy, with $\sigma$ the unit of length. In addition, bonded beads also interact via the Finite-Extensible-Nonlinear-Elastic (FENE) potential, defined as, 
\begin{equation}
    U_{FENE}(r)=-\epsilon k_F {R_0}^2\log{\left[ 1-\left(\frac{r}{R_0\sigma}\right)^2 \right]} \text{ if } r<R_0\sigma,
\end{equation}
with $R_0=1.5$, and $k_F=15$, setting the maximum bond extension and stiffness, respectively. 

After an initial equilibration run of a single microgel, we generate a low-density initial random configuration of $N=54$ ULC microgels by replicating $N$ times the equilibrated one, while avoiding overlaps, i.e. number density of $\rho = N/V = 1.1 \times 10^{-7}$. 
We then perform NVT MD molecular simulations of the multiple microgels  for at least $5 \times 10^{6}\tau$, where $\tau=\sqrt{m\sigma^2/\epsilon}$ corresponds to the time unit. The temperature is fixed to $k_BT/\epsilon=1$, with $k_B$ the Boltzmann constant, and the timestep to $\delta t =0.002\tau$. Next, we gradually compress the system to a target volume $V$, followed by additional equilibration run of at least $5 \times 10^{6}\tau$. This equilibrated compressed configuration is subsequently used as the starting point for the next compression. By repeating this procedure, we systematically explore the full range of densities. For the calculation of observables, each target $V$ is simulated for at least $3\times 10^7 \tau$ after initial equilibration.  All simulations are performed with the LAMMPS package~\cite{thompson2022lammps}.

{To place ULCs in the broader context of microgel behavior, we also perform analogous simulations of additional systems in which we independently vary the key assembly parameters. In particular, } {to explore the role of network connectivity, we simulate ULC microgels with crosslinkers with valence equal to 4, denoted as ULC$_\mathrm{4p}$, while keeping the single-particle number density fixed to $\rho_\mathrm{ULC}$. In addition, to investigate the effect of particle density at high crosslinker concentration, we also consider a system of $54$ microgels with $N_m\sim17000$ monomers per particle, assembled at $\rho_\mathrm{ULC}$ with total crosslinker concentration of $\sim5\%$, denoted as  $c=5\%_\mathrm{ULC}$. Finally, to explore the effect of the crosslinker concentration in standard microgels, we perform simulations of systems of microgels with $c=1.0\%$ and $c=0.2\%$ with same number of monomers $N_m$ and assembled with single-particle density fixed to $\rho_\mathrm{st}\approx0.08$~\cite{marin2025predicting}.}

\subsection*{Structural Analysis}
To estimate the size of the microgel at a given state point, we calculate the average hydrodynamic radius $R_{H}$. To estimate it, we follow the method introduced in Ref.~\cite{del2021two}, where, 
\begin{equation}
    R_H=2\left[\int_0^\infty \frac{1}{\sqrt{(a^2+\theta)(b^2+\theta)(c^2+\theta)}}d\theta\right]^{-1},
    \label{eq:rh}
\end{equation}
with $a$, $b$, and $c$ the principal semiaxes of the gyration tensor associated to the simplices of the surface mesh that encloses the ULC. The surface mesh is constructed using the alpha-shape algorithm implemented in OVITO~\cite{stukowski2009visualization}, with a probe sphere radius $R_{\alpha}=12\sigma$ that controls the resolution of the mesh. Since the ULCs exhibit a highly extended conformation, the choice of $R_{\alpha}$ is a compromise between minimizing spurious voids within the mesh and avoiding an overestimation of the ULC volume. Other probe radii $R_{\alpha}$ lead to similar results.
We also calculate the gyration radius $R_g$ which yield qualitatively similar results, albeit with much smaller values.
However, the microgels are largely non-spherical, so in order to have a meaningful estimate of the suspension packing fraction, we directly rely on the volume of the microgel as obtained from the surface mesh calculation. 
We thus estimate the nominal packing fraction $\zeta$ as follows:
\begin{equation}
\zeta=NV_{\alpha,0}/V,
\label{eq:zeta}
\end{equation}
where $V_{\alpha,0}$ denotes the average volume of a single ULC under dilute conditions and $V$ the volume of the simulation box.
{The same definitions are applied to all types of studied microgels.}

The overall structure of the individual microgels is further characterized by calculating their average density profiles, 
\begin{equation}
    \rho(r)=\left<\sum_{i=1}^N \delta(|\mathbf{r}_i-\mathbf{r}_{cm}|-r)\right>,
    \label{eq:densprof}
\end{equation}
where $\mathbf{r}_{cm}$ corresponds to the position of the center of mass of the microgel, $\mathbf{r}_i$ that of particle $i$, and $<>$ represent a time average. Despite the very low crosslinker concentration in ULC microgels, their density profiles can still be accurately fitted with the fuzzy-sphere model~\cite{stieger2004small}, which for simplicity can be written as,
\begin{equation}
    \rho(r)\sim A  \textrm{ erfc}\left(\frac{r-R_c}{\sqrt{2}\sigma_s}\right),
    \label{eq:fuzzysphere}
\end{equation}
where $A$ is a fit parameter, $R_c$ denotes the radius of the core, and $\sigma_s$ the half-width of the corona shell.

The very soft architecture of the network increases the likelihood of particle overlap at high concentrations. To characterize this effect, we compute the many-body overlap packing fraction by estimating the volume jointly occupied by $n_b=1,2,\ldots, n$ distinct microgels through the intersections of their surface-meshes. 

Additionally, ULCs undergo strong deformation with respect to their spherical shape. In order to quantify these changes, we estimate the shape parameter $S_p$~\cite{marin2023colloidal,del2024numerical}, which is defined as the ratio of the microgel volume $V_{\alpha}$, with respect to the volume of a sphere, $V_s$ having the same surface area $A_{\alpha}$,
\begin{equation}
    S_p=\frac{V_{\alpha}}{V_s}=6\sqrt{\pi}\frac{V_{\alpha}}{A_{\alpha}^{3/2}}.
    \label{eq:nu}
\end{equation}
A value of $S_p=1$ corresponds to a perfect sphere. This quantity is averaged over all microgels in the suspension.  
Furthermore, to assess the influence of the outer chains on particle anisotropy and to gain deeper insights into internal deformations, we also calculate $S_p$ after removing the monomers located at a distance from the center of mass greater than $l_h$. 
Following Ref.~\cite{del2024numerical}, we define $l_h=R_c+\sigma_s$, where $R_c$ and $\sigma_s$ are obtained from the fuzzy-sphere fits of the ULC density profiles under dilute conditions. In our case, the parameters extracted from the fit at low density are $R_c \sim 41.3\sigma$ and $l_h \sim 59.8\sigma$, with $\sigma_s=18.5\sigma$.

Finally, we characterize the structure of the whole suspension by calculating the radial distribution function $g(r)$ between the centers of mass of the microgels.

\subsection*{Dynamics and Mechanical Response}

To characterize the elastic properties of the system, we calculate the bulk modulus of the suspension as the derivative of the pressure $K=-V(\partial P/\partial V)$. In addition, we estimate the single particle bulk modulus $K_p$ following Ref.~\cite{del2024numerical}. 
We simulate the compression of a single microgel by performing $NVT$ simulations upon the addition of a harmonic force $F(r)=-k(r-R)^2$, where $R$ corresponds to the radius of the spherical confinement and $k=10$ to the strength of the force. We vary the value of $R$ and equilibrate the configuration, after that we compute the microgel volume via the surface mesh of the particle and calculate the individual bulk modulus from the fluctuations of such,
\begin{equation}
    Kp=\frac{k_BT\left<V_{\alpha} \right>}{\left<V_{\alpha}^2 \right>-\left<V_{\alpha} \right>^2}.
    \label{eq:bulkmod}
\end{equation}
 To be able to compare with the overall bulk modulus, we map $R$ to the corresponding $\zeta$ by equaling the dilute conditions data.

Finally, to monitor the dynamics of the system we compute  the mean-squared displacement (MSD) of the microgel center of mass as,
\begin{equation}
    \left\langle r^{2}(t) \right\rangle = \frac{1}{N} \left\langle \sum_{i=1}^{N} {\left\lvert {\bf r}_{cm,i} - {\bf r}_{cm,i}(0) \right\rvert}^{2} \right\rangle \, .
    \label{eq:msd}
\end{equation}
When this quantity reaches a diffusive behavior, we also estimate the long-time self-diffusion coefficient as 
\begin{equation}
D= \lim_{t \rightarrow  \infty} (d \left\langle r^{2}(t) \right\rangle/dt)/6.
\end{equation}


\section*{Acknowledgments} We are grateful to the anonymous Reviewers for their helpful and motivating remarks. We thank G. del Monte for useful discussions and help with analysis algorithms. This project was funded by the European Union HORIZON-MSCA-2022-Postdoctoral Fellowships under grant agreement no. 101106848, MGELS. EZ also acknowledges funding from ICSC – Centro Nazionale di Ricerca in High Performance Computing, Big Data and Quantum Computing, funded by European Union – NextGenerationEU - PNRR, Missione 4 Componente 2 Investimento 1.4. We gratefully acknowledge the CINECA award under the ISCRA initiative, for the availability of high-performance computing resources and support.

%

\end{document}